# Enabling Wireless Communications, Energy Harvesting, and Energy Saving by Using a Multimode Smart Nonlinear Circuit (MSNC)

Rasool Keshavarz[1], Dimitrios L. Sounas[2], Saeed Keshavarz[2], and Negin Shariati[1]

*Abstract*— In this paper, a multimode smart nonlinear circuit (MSNC) for wireless communications (Tx and Rx modes) as well as energy harvesting (EH) and power saving is presented. The proposed MSNC is designed at 680 MHz and has three ports, which are connected to an antenna, and T/R (transceiver) and power-saving modules. According to the input/output power level, the proposed MSNC has three modes of operations; Receiving (Rx), power saving and transmitting (Tx), for low (<-25 dBm), mid (>-25 dBm and <0 dBm) and high (>5 dBm) power ranges, respectively. In the power-saving mode, when the received power is greater than the sensitivity of the Rx module, the excess power is directed to the energy harvesting load (power storage), while the receiving direction is still in place. The fact that the proposed MSNC can manage the received power level smartly and without any external control, distinguishes the proposed MSNC from other EH circuits. The proposed MSNC operates within a power range from −50 dBm to +15 dBm, demonstrates an efficiency of more than 60% in the power-saving mode, and has acceptable matching over a large frequency range. The design procedure of the proposed MSNC along with the theoretical, simulation and measurement results are presented in this paper. Good agreement between theory, simulation and measurement results confirms the accuracy of design procedure.

## I. INTRODUCTION

WITHIN the not-so-distant future, highly-developed electronic devices coordinating with communication modules may have the capacity to change our everyday lives [1]. These devices will be implanted into our ordinary objects, sense the physical world, and collect and trade information by means of the internet [2],[3]. In this scenario, our environment is not inactive anymore. It can be capable of communication with humans and transmit valuable data to us, such as environmental conditions (e.g. temperature and humidity), and human blood pressure [4]. This concept based on systems of smart devices is known as the internet of things (IoT) [5]. In terms of wireless sensor networks (WSNs), the number of sensors and electronic devices interconnected through the internet may reach up to 50 billion globally. Thus, one of the primary challenges of utilizing sensor networks is to feed them with constant power supplies [6]-[8]. The most common power supply for remote electronic devices is batteries. Conventional approaches of replacing and recharging batteries to extend the lifetime of electronic devices are impractical and inefficient for WSNs due to large-scale deployment of sensors. Although the advancement of communications and battery technologies has allowed many devices to operate wirelessly, the battery replacement is still required. In turn, this results in high maintenance costs for changing and disposing batteries, which tend to degrade with time [9], [10].

On the other hand, renewable energy sources, such as wind and sun, are not fitted into most of IoT applications, since these sources are primarily inaccessible in indoor situations and their integration with small sensors is challenging [11], [12].

Luckily, newly-emerged green energy sources address the issue of accessibility of renewable energy sources. Radio-Frequency (RF) wireless power transfer (WPT) has incredible potential to deliver continual and adequate power to sensor systems via remote channels during day and night, both indoor and outdoor [13], [14]. Besides, information and power can be transferred simultaneously to receivers through the air, which is known as simultaneous wireless information and power transfer (SWIPT) [15], [16]. In the WPT or SWIPT scenario, transmitter sends the RF energy to the receiver and through a rectenna (rectifying antenna) system, which comprises an antenna and a rectifier, the power is delivered to a supercapacitor or a battery. As well, existing approaches require manual switching for the different functionalities of the system. It is worth mentioning that the overall performance of an RF energy harvesting system is highly dependent on the rectenna efficiency and sensitivity [17]. Moreover, the supplied devices (e.g. self-sustainable sensors ($S^3$)) need to communicate with a base station or other communication nodes to transfer data and receive proper commands. These problems would be avoided by developing a smart system that operates based on the power level, enabling a time-efficient way to automatically distinguish between the three modes from the power level. In the previous works, the authors only concentrated on WPT [18]-[22], and the importance of power saving is neglected, especially when the data is sent from devices to a base station. The other attempt in the literature to cover all three modes of operation, i.e. Rx, Tx, and power saving modes, is SWIPT. The main downside of the proposed SWIPT systems is that separate circuits or manual switching components are used to cover all operational modes [23], [24].

[1]RF and Communication Technologies (RFCT) research laboratory, University of Technology Sydney, Ultimo, NSW 2007, Australia, e-mail: Rasool.Keshavarz@uts.edu.au; Negin.Shariati@uts.edu.au.

[2] College of Engineering, Wayne State University, 5050 Anthony Wayne Dr., Detroit, MI 48202, e-mail: dsounas@wayne.edu.

One way to realize a system sharing one antenna between TX and RX modules is using magnetic or magnetless circulators [24]-[30]. However, circulators are not sensitive to input or output power level, so they are not suitable for the design of a smart system, which is intended to be used in batteryless devices to operate at three modes: Rx, Tx and power saving. To address this issue a circuit is proposed for wireless sensor networks using a circulator connected to Rx antenna, low-threshold rectifier, and high-power rectifier through ports 1, 2, and 3, respectively [27]. Although this work is a good progress in WPT and energy harvesting fields of study, it suffers from having bulky magnetic circulator making the proposed circuit non-integrable with other parts of the Rx/Tx chain.

In this paper, a multimode smart nonlinear circuit (MSNC) is proposed, which is connected between antenna, Tx/Rx and power storage (Fig. 1). According to the input power level, the proposed MSNC connects the antenna port to the proper section (Rx, Tx and power storage), intelligently. Therefore, the proposed MSNC includes three operating modes in terms of input/output power level: low-power, mid-power and high-power. As shown in Fig. 1, the blue, green, and red paths are related to receiving, receiving along with power saving, and transmitting modes. At low and high input/output power levels, the antenna port is connected to the Rx and Tx sections, respectively, through the proposed MSNC. At mid power level specified with the green arrows, the circuit works in power saving mode; since the input power level is greater than the Rx sensitivity, portion of the input power is directed to the energy harvesting section and stored in the supercapacitor/battery while the Rx path is still in place. Therefore, the power bounds of the power-harvesting mode are determined by the RX sensitivity and the TX power.

The contributions of this paper are summarized as follows:
- For the first time, we designed and analyzed a multimode smart nonlinear circuit (MSNC) which automatically directs input\output signal to the desired ports according to the signal level. The proposed highly sensitive MSNC operates over a wide input power range from $-50$ dBm ($0.01\ \mu W$) to $+15$ dBm ($31.6\ mW$) in three different modes: Tx, Rx, and power saving modes.
- From design and economic perspectives, utilizing a large number of components (e.g. antennas, diodes, switches, etc.) to realize individual switching systems for WPT and Rx/Tx operating modes impose additional expenses, and increase the total footprint. This problem is avoided with the proposed MSNC, making it compact and low cost, and with a potential to be integrated with various industrial sensors and IoT systems.
- The communication link is always active even in power saving mode. This eliminates any discontinuity in the data link, which is of paramount importance in Wireless Sensor Networks (WSNs).
- The design guide procedure for arbitrary load impedance and operational frequency bands is presented and equations are derived. Therefore, the design procedure can be applied to the desired frequency and load impedances.

The proposed MSNC is insensitive to phase mismatch in multi-tone excitations, making it a good candidate for real WPT scenarios.

The organization of this paper is as follows: Step by step design guide and equations of the proposed multimode smart nonlinear circuit (MSNC) topology are presented in section II. The MSNC performance is validated by analytical, simulation, and measurement results in section III. Finally, conclusions are provided in section IV.

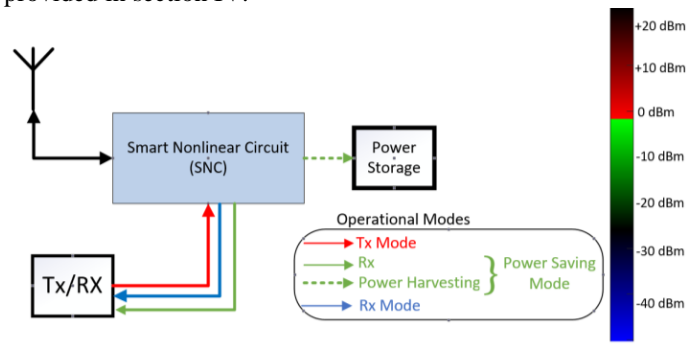

Fig. 1. Proposed multimode smart nonlinear circuit (MSNC) conceptual diagram and its performance within different power ranges.

## II. THEORY AND DESIGN PRINCIPLE

The proposed multimode smart nonlinear circuit (MSNC) is shown in Fig. 1, and it is designed to have three ports, specifically Ant (antenna-P1), TR (Tx/Rx-P2), and PS (power saving-P3) ports. Ports Ant and TR are connected to an antenna and a Transceiver (Tx/Rx), respectively, and the DC power is collected at the load resistance of the PS port. The detailed schematic of the proposed MSNC is depicted in Fig. 2 and it consists of a branch-line coupler (BLC), anti-parallel Schottky diodes, matching networks (MN1 and MN2) and rectifiers. Green, red and blue arrows in Fig. 1 and Fig. 2 illustrate the wave directions in the power saving, Tx and Rx modes, respectively. Therefore, according to the input/output power level, the proposed multimode MSNC switches between three modes: Rx, Tx and power saving. To enable switching between the three operational modes, the proposed MSNC includes parallel and series configurations of Schottky diodes, taking advantage of the variation of the Schottky diodes impedances against the input power level. When it comes to the BLC, it consists of microstrip transmission lines with characteristic impedances $Z_0/\sqrt{2}$ and $Z_0$ for the horizontal and vertical lines, respectively. The length of the lines of the BLC is $\lambda_g/4$, where $\lambda_g$ is effective wavelength.

The circuit should be sensitive to the input power level, in other words, it should have different modes of operation according to the input power level. This can be achieved by adding non-linear loads creating various impedances depending on the input power at nodes $a_2$ and $a_3$ in Fig. 2. Specifically, anti-parallel Schottky diodes are used between $b_2$ and $c_2$ as well as $b_3$ and $c_3$ to create non-linear impedances. Moreover, the series diodes and the shunt capacitors of the RF to DC rectifier circuits are connected between $d_2$ and $e_2$ and $d_3$ and $e_3$, before connecting the non-linear impedances between the nodes $a_2$-$b_2$, $a_3$-$b_3$, $c_2$-$d_2$, and $c_3$-$d_3$.

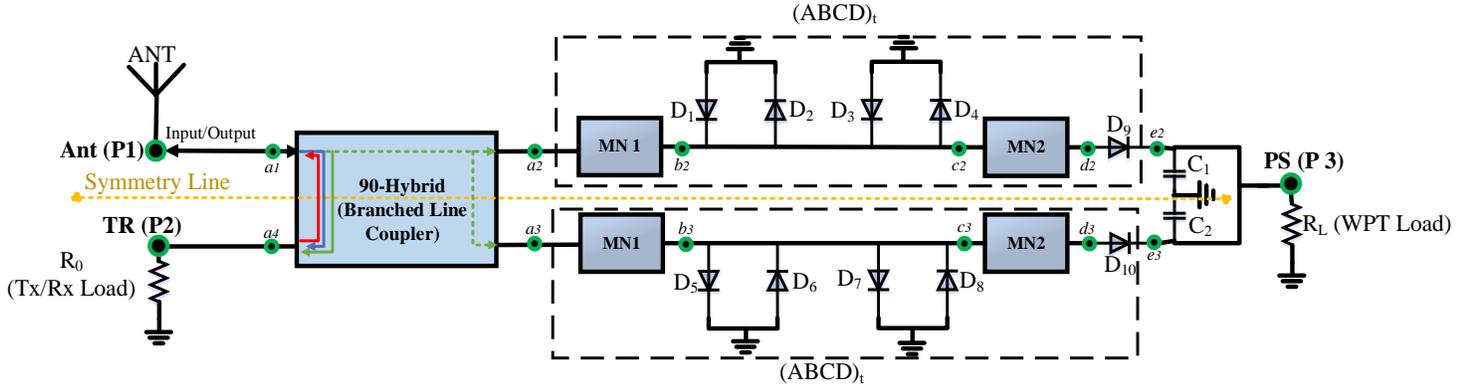

Fig. 2. Schematic of the proposed MSNC which consists of branched-line coupler, rectifiers, parallel diodes, matching networks, WPT and Tx/Rx loads and its performance as a multimode smart nonlinear circuit.

It is worth noting that the same matching networks are used in both the top and bottom paths to create symmetry in the proposed circuit. Furthermore, the impedances of port Ant (antenna) and TR (Tx/Rx) are the same (50 Ω).

To analyse the proposed circuit, the nonlinear branches on the left side of the network are modelled with equivalent nonlinear impedances $Z_{a2}$ and $Z_{a3}$ connected to $a_2$ and $a_3$ (Fig. 3a). Due to the symmetry of the circuit, $Z_{a2}$ and $Z_{a3}$ have the same values. As regards to the input power, three modes of operation can be defined for the proposed circuit, specifically receiving (Rx), power saving, and transmitting (Tx) modes for low power (Rx), mid power (power saving), and high power (Tx) levels, respectively. Receivers and transmitters operate with low power and high powers, respectively, so that the proposed circuit can operate within the power ranges dictated by the RX sensitivity and TX power. Likewise, the circuit operates in the power saving mode in the mid power case. The design guide of the proposed circuit is presented in the following paragraphs.

There are several degrees of freedom in the design of the proposed circuit, including the BLC, the matching networks, the non-linear diodes and the rectifiers. Here, we assume all sections of the circuit are determined (BLC, diodes and capacitors), but matching networks (MN1 and MN2) connected to ports 2 and 3 of BLC, so the circuit design is restricted to the design of the matching networks. Owing to the symmetry of the circuit, an even-odd-mode analysis can be used. Equivalent circuits of the odd and even modes are presented in Fig. 3b and Fig. 3c, respectively. As can be seen in Fig. 3b and Fig. 3c, the middle of the vertical branches of the BLC are grounded and open circuit for the odd and even modes, respectively.

According to [30], the signals at each port of the BLC can be expressed as:

$$A_1 = \tfrac{1}{2}(\Gamma_e + \Gamma_o) \tag{1a}$$
$$A_2 = \tfrac{1}{2}(T_e + T_o) \tag{1b}$$
$$A_3 = \tfrac{1}{2}(T_e - T_o) \tag{1c}$$
$$A_4 = \tfrac{1}{2}(\Gamma_e - \Gamma_o) \tag{1d}$$

where $\Gamma_e, \Gamma_o, T_e,$ and $T_o$ are the even- and odd-mode reflection and transmission coefficients at the BLC ports, which are given by

$$\Gamma_i = \frac{(B_{BLC,i} - C_{BLC,i} Z_S Z_{A,i}) + (A_{BLC,i} Z_{A,i} - D_{BLC,i} Z_S)}{(B_{BLC,i} + C_{BLC,i} Z_S Z_{A,i}) + (A_{BLC,i} Z_{A,i} + D_{BLC,i} Z_{A,i})} \tag{2a}$$

$$T_i = \frac{2 Z_{A,i}}{(B_{BLC,i} + C_{BLC,i} Z_S Z_{A,i}) + (A_{BLC,i} Z_{A,i} + D_{BLC,i} Z_{A,i})} \tag{2b}$$

$i = e, o$

where $A_{BLC,i}, B_{BLC,i}, C_{BLC,i}$ and $D_{BLC,i}$ are the odd- and even-mode ABCD matrix parameters of the BLC in Fig. 3b and Fig. 3c.

$$(ABCD)_{BLC,e} = \tfrac{1}{\sqrt{2}}\begin{bmatrix} -1 & jZ_0 \\ j/Z_0 & -1 \end{bmatrix} \tag{3a}$$

$$(ABCD)_{BLC,o} = \tfrac{1}{\sqrt{2}}\begin{bmatrix} 1 & jZ_0 \\ j/Z_0 & 1 \end{bmatrix} \tag{3b}$$

$Z_{A,i}$ is the load impedance in the even- ($Z_{Ae}$) or odd-mode - ($Z_{Ao}$) for nodes $a_2$ and $a_3$, and $Z_S$ is the source impedance (antenna or T/R impedances), which is considered 50 ohms.

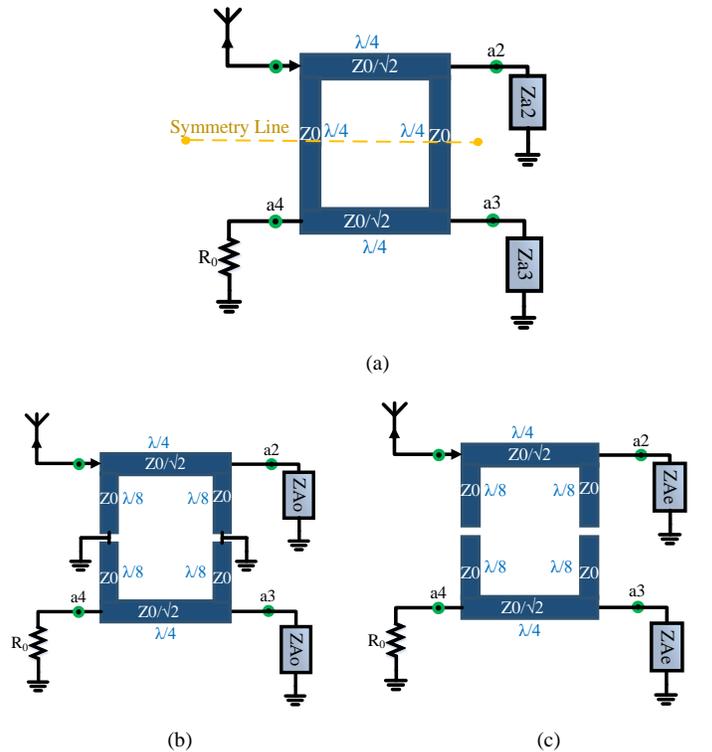

Fig. 3. Equivalent circuits of odd and even modes of the BLC a) BLC with nonlinear loads (proposed MSNC) where $Z_{a2} = Z_{a3}$, b) odd, and c) even modes.

The design procedure aims to manipulate the amplitude expressions in Eq. 1 in a way that we find conditions that end up in different operation modes (Tx, Rx and power saving). Due to the $90^0$ phase difference between the output ports of BLC coupler ($a_2$ and $a_3$ in Fig. 2), the output signals $A_2$ and $A_3$ can be defined as follows:

$$A_2 = -jk/\sqrt{2} \quad (4a)$$
$$A_3 = -k/\sqrt{2} \quad (4b)$$

where $k^2/2$ is the power-transfer power ratio between the output ports of the BLC, i.e. nodes $a_2$ and $a_3$. Moreover, $k$ can be used to control the transmission coupling between ports 1 and 2.

According to Eqs. 1(b), 1(c) and 2(b), the only parameters that are not pre-determined for are $Z_{Ae}$ and $Z_{Ao}$. Our goal is to determine the dependence between these impedances and $k$, to find the impedances that are needed to be connected to the output ports of the BLC to achieve a given power-transfer ratio $k$ and therefore a given coupling between the ANT and TX/RX ports. This dependence can be calculated numerically using MATLAB. Fig. 4a and 4b show the real and imaginary parts of $Z_{Ae}$ and $Z_{Ao}$ against different values of $k$, respectively. According to this figure, the real parts of $Z_{Ae}$ and $Z_{Ao}$ are 50 Ω when all input power is delivered to the rectifiers ($k=1$).

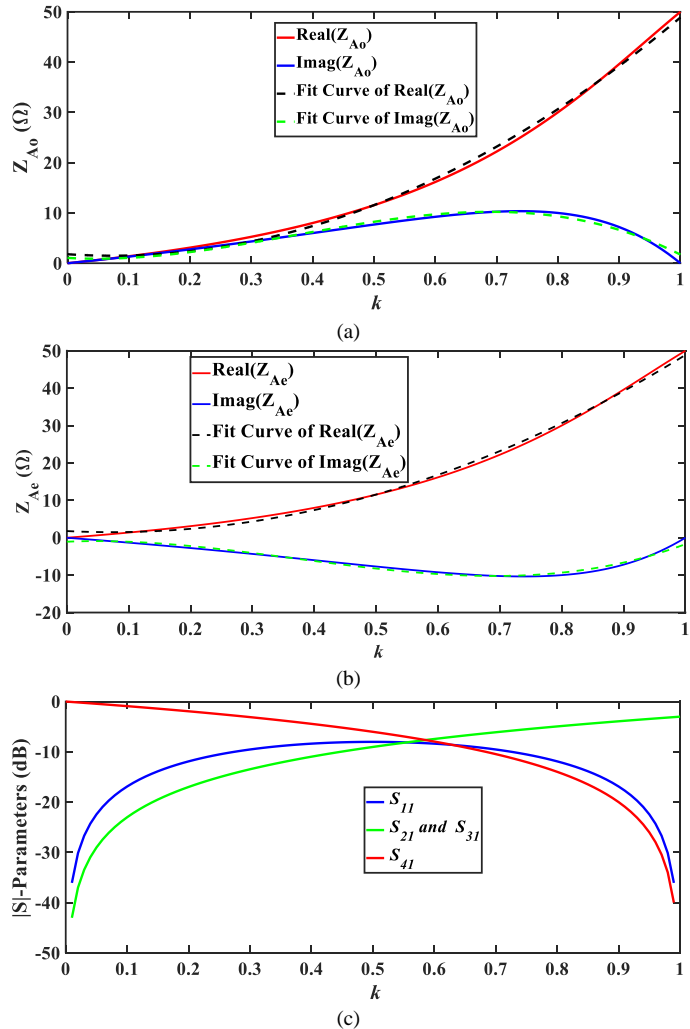

**Fig. 4**. a) $Z_{Ao}$, b) $Z_{Ae}$, and c) S-parameters of BLC versus k.

To show the MSNC performance, the S-parameters of the BLC versus $k$ are plotted in Fig. 4c for the $Z_{Ao}$ and $Z_{Ae}$ determined from the previous analysis (Eqs. 2(a), 1(a) and 1(d)). As can be seen from this figure, $|S_{21}|$ and $|S_{31}|$ of the BLC increase from very low values to 3 dB as $k$ increases from 0 to 1, and $|S_{41}|$ decreases from 0 dB to low values as $k$ increases from 0 to 1. This behavior indicates that at lower values of $k$, most of the power is delivered to node $a_4$, where we have T/R modules, and at higher values of $k$, most portion of the input power is delivered to nodes $a_2$ and $a_3$ equally to be stored at PS port. Also, the BLC is well matched ($|S_{11}|<-10$ dB) for all values of $k$. This analysis shows that if the circuit is designed to operate with a low $k$ at the low and high-power ranges, and with a high $k$ at the mid-power range, it can act as a smart circuit, which is sensitive to input power level and is able to switch between three modes (Tx, Rx, and power saving), automatically.

To make the design of the circuit more efficient, we found closed form equations for the real and imaginary parts of $Z_{Ao}$ and $Z_{Ae}$ versus $k$ by using the curve fitting tool in MATLAB. These equations are

$$Real(\overline{Z}_{Ae}) = Real(\overline{Z}_{Ao}) = 55.1k^2 - 8k + 1.8 \quad (5)$$
$$Imag(\overline{Z}_{Ae}) = -Imag(\overline{Z}_{Ao}) = 70k^3 - 77.6k^2 + 6.9k - 1.1 \quad (6)$$

As it can be seen in Fig. 4, $\overline{Z}_{Ae}$ and $\overline{Z}_{Ao}$ are in good agreement with $Z_{Ae}$ and $Z_{Ao}$.

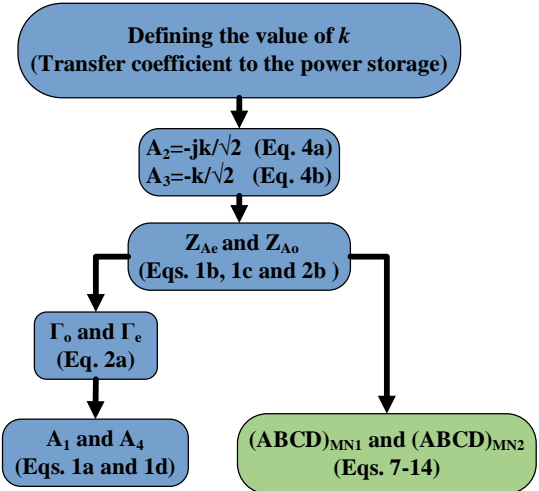

Fig. 5. Design procedure of the proposed MSNC.

According to the aforementioned points, the design procedure of the proposed MSNC is shown in Fig. 5. As shown in this figure, the desired value of $k$ should be determined at the first step, then the equivalent amplitudes transferred to nodes $a_2$ and $a_3$, i.e. $A_2$ and $A_3$ in Eq. 4a and Eq. 4b should be calculated. By manipulating Eq. 1b, Eq. 1c, and Eq. 2b, the values of $Z_{Ae}$ and $Z_{Ao}$ can be determined. Moreover, $A_1$ and $A_4$ are calculated using Eqs. 2a, 1a and 1d.

Next step in the design procedure is the design of $Z_{Ae}$ and $Z_{Ao}$ by using matching networks (MN1 and MN2). Provided that, we can implement $Z_{Ae}$ and $Z_{Ao}$ as Eq. 5 and Eq. 6 using MN1 and MN2, the proposed MSNC can operate as a smart circuit to switch between Tx, Rx and power saving modes regarding the power level, automatically. According to Fig. 2, $Z_{Ae}$ and $Z_{Ao}$ are calculated as

$$Z_{A,i} = \frac{A_t Z_{Li} + B_t}{C_t Z_{Li} + D_t} \quad (i=e,o) \tag{7}$$

where $Z_{Li}$ is the impedance at $d_2$ and $d_3$ in the even- and odd-mode with respect to the symmetry line in Fig. 2, given by

$$Z_{Le} = 1/(2R_L) + j\omega C \tag{8a}$$
$$Z_{Lo} = 0 \tag{8b}$$

$A_t$, $B_t$, $C_t$, and $D_t$ in Eq. 7 are the parameters of *ABCD* matrix of the loading chain at ports 2 and 3 (between *a* and *e* points in Fig. 2) and can be calculated as

$$(ABCD)_t = (ABCD)_{MN1} \times (ABCD)_{PD} \times (ABCD)_{MN2} \times (ABCD)_{SD} \tag{9}$$

where $(ABCD)_{MN1}$, $(ABCD)_{PD}$, $(ABCD)_{MN2}$, $(ABCD)_{SD}$, and $(ABCD)_t$ are the *ABCD* matrices of the first matching network, the parallel diodes (acting as non-linear impedances), the second matching network, the series diode of the rectifier circuit, and the total load connected to nodes $a_2$ and $a_3$, respectively. The expressions of $(ABCD)_{PD}$ and $(ABCD)_{SD}$ are given by

$$(ABCD)_{PD} = \begin{bmatrix} 1 & 0 \\ 4Y_D & 1 \end{bmatrix} \tag{10a}$$
$$(ABCD)_{SD} = \begin{bmatrix} 1 & Z_D \\ 0 & 1 \end{bmatrix} \tag{10b}$$

where $Z_D$ and $Y_D$ are the equivalent nonlinear impedance and admittance of a Schottky diode, respectively, and their values depend on the diode voltages ($v_D$). According to Fig. 2, the diode voltages for four parallel diodes and series diode are $v_c$ and $v_{(d-e)}$, respectively. Furthermore, due to the symmetry of the circuit and the choice of the same diodes for D1, D2, D3, and D4, all of them have the same admittances as in Eq. 11:

$$Y_{D1} = Y_{D2} = Y_{D3} = Y_{D4} = Y_D \tag{11}$$

Moreover, a nonlinear model of the Schottky diode is considered, including a nonlinear conductance $G(v_D)$ in parallel with a nonlinear capacitance $C(v_D)$ and a fixed resistance $R_s$ (Fig. 6). Hence, the Schottky diode is represented by nonlinear equation as shown in Eq. 12.

$$I_D = I_s \exp(\alpha v_D) - \frac{C_0}{\sqrt{1 - \frac{v_D}{V_0}}} \frac{dv_D}{dt} \tag{12}$$

where $C_0$ is the static capacitance of the Schottky set by DC bias $V_{DC}$, $I_s$ is the saturation current, $\alpha = \frac{1}{nV_T}$ ($V_T = \frac{k_b T}{q}$ is the thermal voltage, $n$ is the diode ideality factor, $k_b$ is the Boltzman constant, $T$ is the junction temperature, and $q$ is the electron charge).

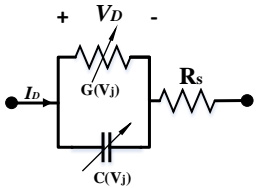

**Fig. 6.** Equivalent circuit model of a Schottky diode [13].

The impedances of the parallel and series diodes $Y_D$ and $Z_D$ in Eq. 10 are functions of the voltage and current of each diode. Fig. 7 plots the real and imaginary parts of these impedances versus the diode voltage and current. The imaginary part of diode impedance is purely capacitive and this capacitance in the equivalent circuit model increases at low voltage range. Moreover, the total impedance of diode approximates short-circuit and open-circuit, at high and low voltages, respectively. A similar response is expected to happen for $Z_{Ae}$ and $Z_{Ao}$. The sharp variation of diode impedance and admittance against the voltage and current for series and parallel diodes, respectively, enables the design of a smart nonlinear circuit which can automatically change its performance with different power levels of the excitation signal. Then, the main purpose of the design is to determine matching networks that lead to desired values of $Z_{Ae}$ and $Z_{Ao}$ depending on the input power level, according to Eqs. 5 and 6. Specifically, the MNs need to be designed so that for low and high power levels $Z_{Ae}$ and $Z_{Ao}$ take values from the portion of Fig. 4(a) and (b) over low *k*-values, while for mid-power levels, $Z_{Ae}$ and $Z_{Ao}$ take values from the portion of Fig. 4(a) and (b) over high *k*-values. This can be done as follows. From Eqs. 5 and 6, we can determine $A_t, B_t, C_t$ and $D_t$ as functions of *k*. Furthermore, from Fig. 7 we know $Z_D$ and $Y_D$ as functions of $V_D$ and $I_D$, which for a given power level and *k*, are functions of the matching networks and *k*. This approaches leads to the equation

$$(ABCD)_t(k) = (ABCD)_{MN1} \times (ABCD)_{PD}(k, (ABCD)_{MN1}, (ABCD)_{MN2}) \times (ABCD)_{MN2} \times (ABCD)_{SD}(k, (ABCD)_{MN1}, (ABCD)_{MN2}) \tag{13}$$

which is implicit in $(ABCD)_{MN1}, (ABCD)_{MN2}$.

The *ABCD* matrices of the matching networks are unknowns, so the parameters of them which are shown in Eq. 14 should be calculated.

$$(ABCD)_{MN1} = \begin{bmatrix} A_{MN1} & B_{MN1} \\ C_{MN1} & D_{MN1} \end{bmatrix} \tag{14a}$$

$$(ABCD)_{MN2} = \begin{bmatrix} A_{MN2} & B_{MN2} \\ C_{MN2} & D_{MN2} \end{bmatrix} \tag{14b}$$

According to Eq. 9, $(ABCD)_t$ equals to Eq. 15.

$$(ABCD)_t = \begin{bmatrix} A_{MN1} & B_{MN1} \\ C_{MN1} & D_{MN1} \end{bmatrix} \cdot \begin{bmatrix} 1 & 0 \\ 4Y_D & 1 \end{bmatrix} \cdot \begin{bmatrix} A_{MN2} & B_{MN2} \\ C_{MN2} & D_{MN2} \end{bmatrix} \cdot \begin{bmatrix} 1 & Z_D \\ 0 & 1 \end{bmatrix} \tag{15}$$

Due to the complexity of solving nonlinear equation (Eq. 10) to find matching network parameters, here we use Genetic algorithm, which is available in Advanced Design System (ADS) software to derive $(ABCD)_{MN1}$ and $(ABCD)_{MN2}$. Genetic algorithms provide a direct search optimization method, by using trial parameter sets that are allowed to evolve towards a set that gives progressively better performance. In this optimization method, the dimensions of each part of matching circuits are considered as the variables, while the goal is to find $(ABCD)_{MN1}$ and $(ABCD)_{MN2}$ components in such a way that Eq. 10 is met. The simulation and measurement results are presented in Section III.

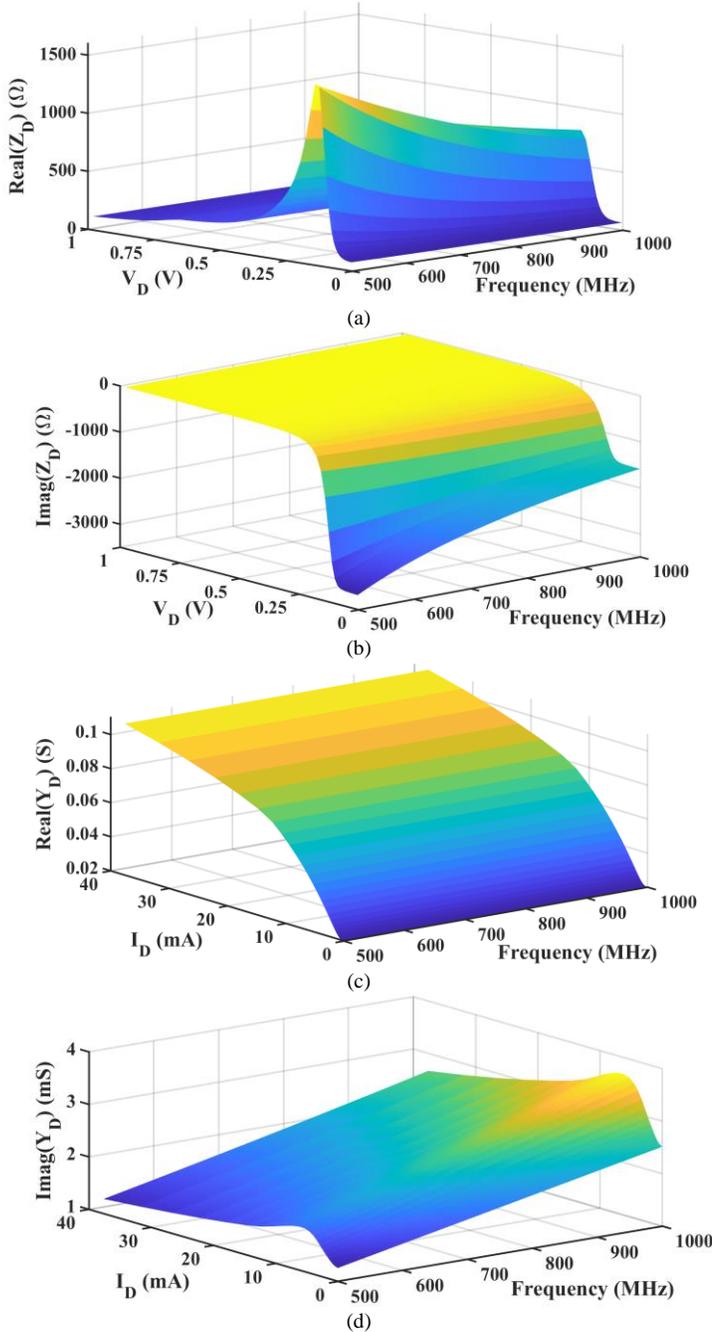

Fig. 7. Real and imaginary parts of parallel and series diodes versus diode voltage and current: a) real, b) imaginary parts of series diode impedance ($Z_D$), c) real, d) imaginary parts of four parallel diodes admittance ($Y_D$), against frequency (550 to 950 MHz) and input/output power level (-40 to 20 dBm).

## III. SIMULATION, MEASUREMENT AND DISCUSSION

To verify the functionality of the proposed multimode MSNC concept, a microstrip prototype operating at 680 MHz has been designed, fabricated, and measured (Fig. 8). Following the MSNC schematic in Fig. 1(a), the layout is developed and a photograph of the fabricated prototype is shown in Fig. 8a. The prototype is fabricated on a Rogers RO4003C substrate with a relative dielectric permittivity 3.55, a 1.52 mm dielectric thickness, a 0.0027 dielectric loss tangent, and a 35 μm metal thickness.

Moreover, SMS7621 from Skyworks, a low-barrier and zero-bias Schottky diode, is used for the series and parallel diodes in Fig. 2 ($D_i$, $i = 1, ... ,10$). The load resistance ($R_L$) is 11 KΩ and the rectifier capacitances ($C_1$ and $C_2$) are 100 nF. The Spice parameters of the diode are presented in Table I. The measurement setup of the proposed MSNC is shown in Fig. 8b. The equipment used to perform the measurement is listed in Table II.

In this section, the simulation and measurement results of the proposed MSNC along with the discussion on the achieved results are presented to verify the functionality of the MSNC as a 3-mode circuit with regards to the input power level.
level.

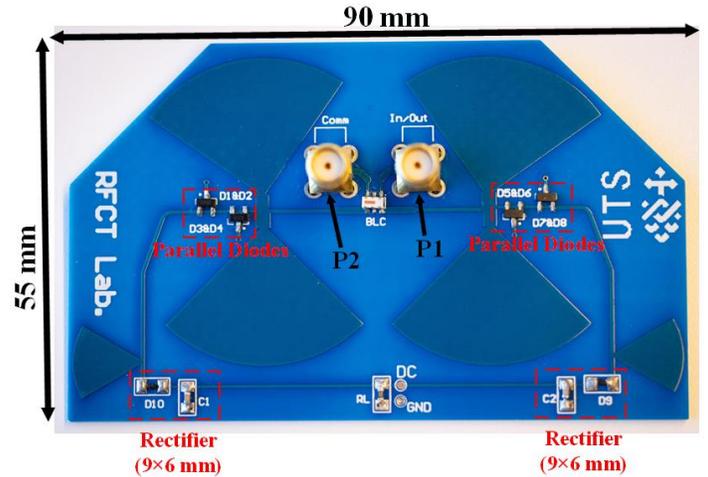
(a)
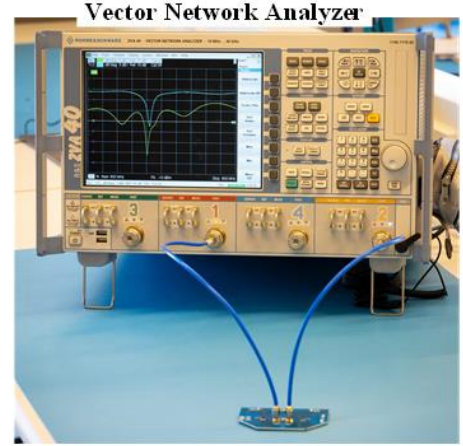
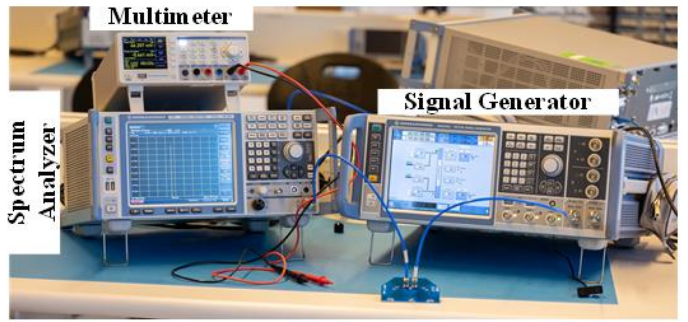
(b)

Fig. 8. a) Multimode smart nonlinear circuit (MSNC) fabricated prototype, and b) measurement setup.

## A. Matching Networks

As mentioned in Section II, the proposed matching networks (MN1 and MN2) are designed based on Genetic optimization algorithm using ADS software at 680 MHz. We specified the number of desired maximum trial/iteration, 1000, during the optimization process. For random optimizers like Genetic algorithm this value represents the number of trials to attempt. In the optimization procedure, we had some optimization variables like length and width of the transmission lines and stubs (Table III). The goal in the optimization process was |S11|<-10 dB in the desired frequency range which reflects good matching condition. After running the optimization algorithm and achieving the final goal, the parameters of the transmission lines and stubs are extracted. Two radial stubs are designed to realize MN1 and MN2 (Fig. 9a) and the dimensions are presented in Table III. To demonstrate that the proposed MSNC operates as a 3-mode (Tx, Rx and power saving) nonlinear circuit, $Z_{Ae}$ and $Z_{Ao}$ of the implemented MSNC using MN1 and MN2 should be the same as the calculated values in Fig. 4a and Fig. 4b, or Eq. 5 and Eq. 6. The simulated results of $Z_{Ae}$ and $Z_{Ao}$ are presented in Fig. 9b and Fig. 9c, considering double radial stubs as MN1 and MN2. These plots are derived by sweeping the input power and calculating $k$ for each power level. According to these figures, there is a good agreement between the theory and simulation values of even and odd mode impedances. The behavior of the simulated curves of $Z_{Ae}$ and $Z_{Ao}$ is as expected, so it is promising that the proposed MSNC operates as a 3-mode nonlinear network which can automatically switch between operational modes based on the input power.

TABLE I
SPICE PARAMETERS OF SCHOTTKY DIODE SMS7621

| $B_V(V)$ | $C_{J0}(pF)$ | $E_G(eV)$ | $I_{BV}(A)$ | $I_S(A)$ | $N$ | $V_J(V)$ | $R_s(\Omega)$ |
|---|---|---|---|---|---|---|---|
| 3 | 0.1 | 0.69 | 1×10E-5 | 4×10E-8 | 1.05 | 0.51 | 12 |

TABLE II
USED EQUIPMENT IN THE MEASUREMENT SETUP

| Vector Network Analyzer | Signal Generator | Spectrum Analyzer | Multi Meter |
|---|---|---|---|
| VNA-ZVA40 | SMW200A | FSV3000 | HMC8012 |

TABLE III
DIMENSIONS OF MATCHING NETWORKS MN1 AND MN2

| Matching Networks | W (mm) | L (mm) | $r_1$ (mm) | $r_2$ (mm) | $\alpha_1$ (Deg) | $\alpha_2$ (Deg) |
|---|---|---|---|---|---|---|
| MN1 | 4.9 | 2.5 | 18.5 | 19.5 | 89 | 90 |
| MN2 | 0.5 | 4.5 | 9.7 | 0 | 60 | 0 |

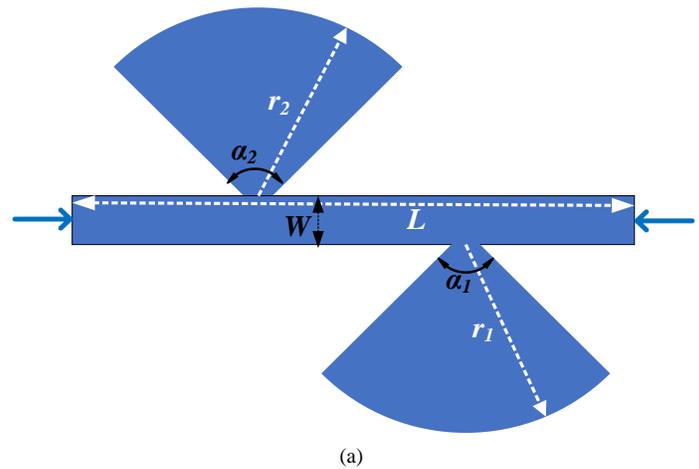

(a)

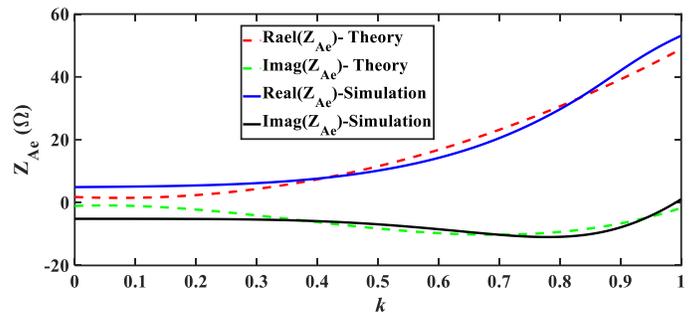

(b)

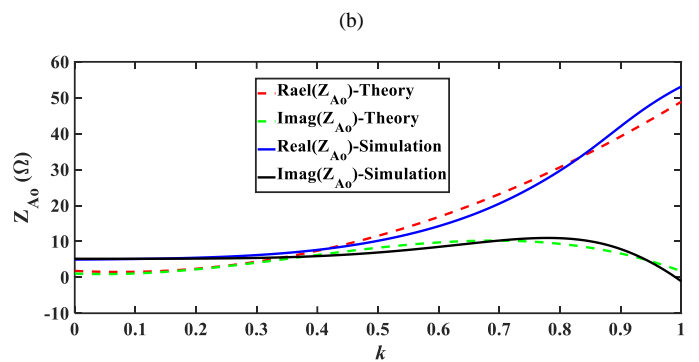

(c)

Fig. 9. a) Schematic of double radial stubs matching network (MN1 and MN2), simulated results of, b) $Z_{Ae}$ and, c) $Z_{Ao}$, considering matching networks.

## B. S-parameters (Return loss and Tx/Rx mode)

As mentioned in the previous section, the behavior of simulated $Z_{Ae}$ and $Z_{Ao}$ by using the designed MN1 and MN2, is the same as the theoretical one. Hence, it is expected that the proposed MSNC with the designed MN1 and MN2 shown in Fig. 9a achieves the desired functionality. In this section, the S-parameters of the proposed MSNC are presented to verify the 3-mode functionality of the circuit. According to the results of the theoretical section (Fig. 4c), the proposed MSNC should exhibit good matching condition for all input power levels; which are low, mid and high. Fig. 10a demonstrates the simulated |S11| and |S22| of the MSNC versus frequency and power, which are the same due to the symmetry of MSNC network. |S11| and |S22| are matched over a broad input power range from $-50$ to $+15\ dBm$. Due to the non-linearity of the diode impedance with respect to input power and frequency, the input matching changes over with input power and frequency. |S12| and |S21| of the

proposed MSNC are presented in Fig. 10b and due to the symmetry of MSNC circuit, they are the same. As shown in this figure, the level of the transmission coefficient at 680 MHz for the mid power range has a null, meaning that for that power level the power is delivered to the rectifier load ($R_L$) and the MSNC is working in the power saving mode. For other power ranges (low and high ranges) at 680 MHz, the $|S_{12}|$ and $|S_{21}|$ have high values (>−1 dB), showing that for such power levels, which are the levels of the Tx and Rx modes, the circuit exhibits a low loss path between port 1 and 2.

Fig. 11 depicts the measured S-parameters of the proposed MSNC. The S-parameters were measured using a Vector Network Analyzer (VNA-ZVA40), which was re-calibrated for each input power level. In Fig. 11, the measured $|S_{11}|$ and $|S_{22}|$ of the MSNC are presented for different input powers ($P_{in} = -40\ to +15\ dBm$) at 680 MHz. According to Fig. 11a, the return loss at port 1 or 2 is better than 13 dB at 680 MHz for the entire power range. Moreover, the minimum in the transmission coefficients between ports 1 and 2 ($|S_{12}|$ and $|S_{21}|$) occurs at −5 dBm, which shows that the power transfer to the power saving load (Port 3 in Fig. 1 and 2) is maximized around this power range. Fig. 12 shows the measured output power at port 2 against the input power at port 1 at 680 MHz in comparison with a linear response. The green zone shows the regione of the power saving mode, and it includes the null in the transmission power between ports 1 and 2. The input power within this null area is delivered to the power saving load (port 3) of the proposed MSNC. Moreover, the blue and red zones illustrate the Rx and Tx mode, respectively.

The measured and simulation results are in good agreement to each other, with any differences attributed to the uncertainties in the diode characteristics and manufacturing tolerances. The clear correlation between measurement and simulation results validates the accuracy of design procedure and theoretical analyses of the proposed MSNC presented in section II.

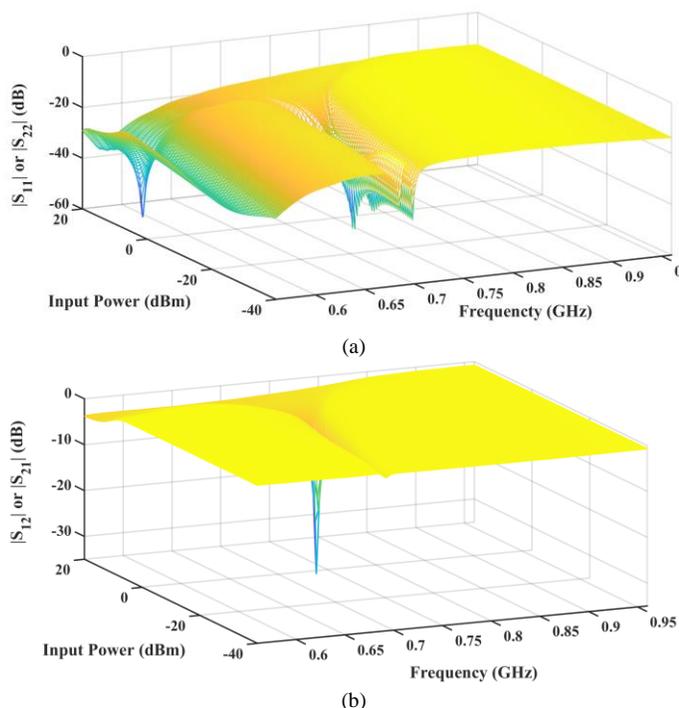

Fig. 10. Simulated S parameters of the proposed MSNC versus frequency (550 to 950 MHz) and input/output power level (-40 to 20 dBm), a) $|S_{11}|$ or $|S_{22}|$, b) $|S_{12}|$ or $|S_{21}|$.

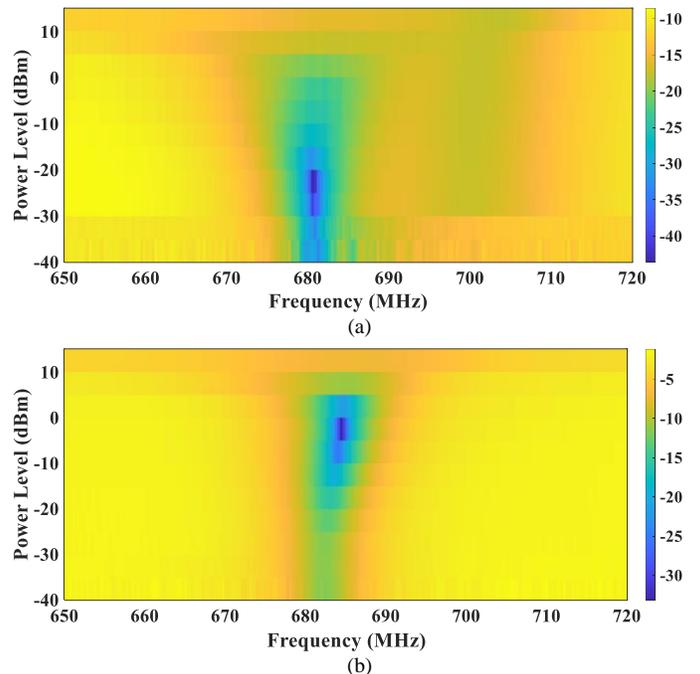

Fig. 11. Measured S parameters of the MSNC over frequency and input/output power level, a) $|S_{11}|$ or $|S_{22}|$, b) $|S_{12}|$ or $|S_{21}|$.

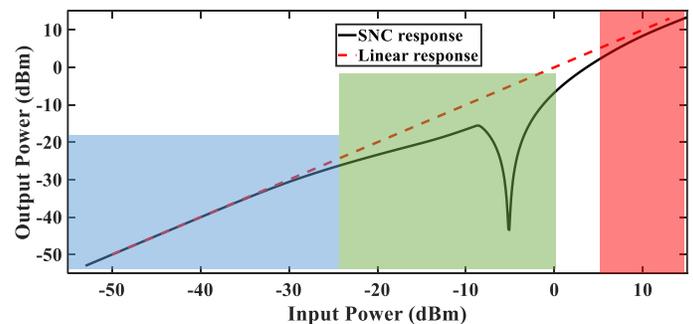

Fig. 12. Power transmission between ports 1 and 2 of the MSNC for different power levels in Rx (blue), power saving (green) and Tx (red) modes compared with the linear response.

### C. Rectifier Performance

The S-parameter analysis presented in the previous section describes the performance between the RF ports of the proposed MSNC, i.e., ports P1 and P2. On the other hand, the output signal of the other port, P3, is in the DC regime, as it is connected to the output of the rectifier circuit. In the power saving mode, RF energy which is directed to ports 2 and 3 of the BLC, is converted to a DC voltage through the rectifiers. These rectifiers consist of series Schottky diodes (SMS7621) and parallel capacitors ($C_1$, $C_2$), as in Fig. 2. Hence, to investigate the power transmission efficiency to port P3, we need to look at the DC power at this port. Simulation and measurement results of the output DC power and the RF to DC conversion efficiency of the proposed MSNC are presented in Fig. 13. Measurements were performed using PSG Vector Signal Generators (SMW200A) as RF power sources, and recording of the output DC voltage across the load resistance was achieved with a Fluke digital voltmeter.

Efficiency is one of the key metrics of a WPT system. According to Fig. 2, the overall efficiency for the WPT part of the MSNC can be defined as

$$\eta_t = \frac{P_{DC}}{P_{in}} \tag{16}$$

Where $P_{DC}$ is the output DC power at port 3 ($R_L$) and $P_{in}$ is the input power of MSNC at port 1 (antenna port). According to Eq. 16, Fig. 13b presents the overall efficiency of the MSNC circuit in WPT mode. The efficiency is over 60% for $P_{in} = -15$ to $-5\ dBm$ (31.6 to 316 $\mu W$) and since the MSNC consists of two parallel rectifiers, the saturation point of the output DC power against the input power level is higher than a single rectifier, leading to high efficiency over a wide input power range. It is worth noticing that the efficiency exhibits a peak at -10 dBm, which is within the power-saving mode regime.

The analysis to this point has been developed assuming that the input signal consists of a single tone. On the other hand, a realistic scenario is expected to involve broadband signals. To investigate how well our system performs in such cases, we analyzed its response under excitation with multiple tones from its input. Hence, a set of concurrent multi-tone signals with the same phase is applied to the MSNC and simulation results are presented in Fig. 14. As can be seen, the output DC power of the proposed MSNC are compared for 1, 3 and 5 tones. The generated output DC power of the sensitive MSNC is saturated by applying 3- and 5-tone excitation at the low input power of 60 $\mu W$ and 130 $\mu W$, respectively. As the number of RF sources increases, the generated stacked output DC power increases, resulting in a reduction of the saturation level (knee point in Fig. 14). It means, the rectifier provides maximum DC power (170 μW) in the output while the level of incoming tones is low but the number of tones increases.

In a realistic scenario, the tones that make a broadband signal are expected to have random phases. To study this case, Fig. 15 shows simulation results of the output DC power for 3-tone excitation signal when there is a phase mismatch between applied tones. In this scenario, the phase mismatch between adjacent tones are considered 0°, 22.5°, 45°, 67.5°, and 90°. It can be seen that there is not significant difference between the results for different values of phase mismatch. Therefore, the proposed MSNC is insensitive to the phase mismatch for multi-tone excitation, and so is a good candidate for real WPT scenarios.

A comparison between our proposed MSNC and some other related designs is given in Table IV. As can be seen, the RF-to-dc power conversion efficiency of the proposed MSNC is comparable with the other works at low power ranges while our proposed rectifier is single band [31-35]. Moreover, the proposed MSNC is not only a rectifier but also it can operate in three operational modes regarding the input power level. A 2-mode rectifier is proposed in [36] for low (EH) and high (WPT) power ranges. This system includes circulator which is bulky at low frequency ranges. As mentioned in Section II, the proposed MSNC is implemented using Schottky diodes and branched-line coupler and this configuration is suitable for low frequency applications.

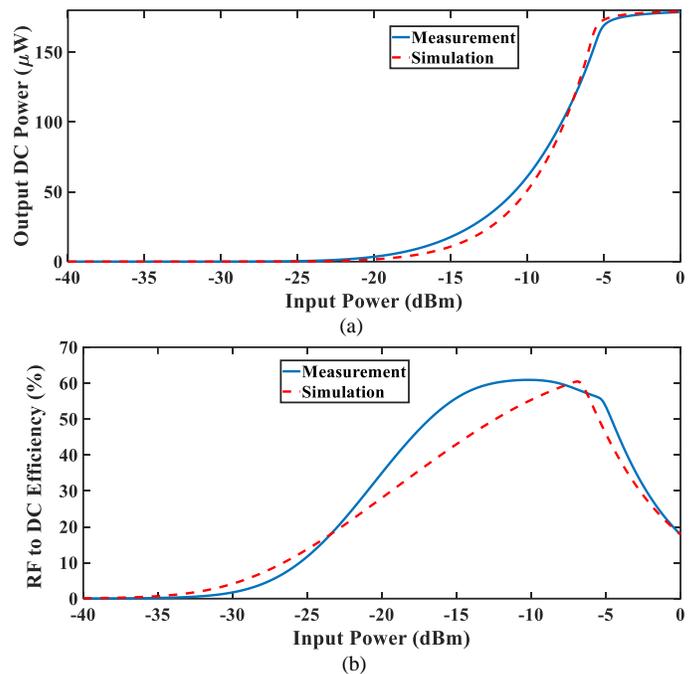

Fig. 13. Simulated and measured results of the proposed MSNC in power saving mode over a broad input power range of $-40\ to\ 0\ dBm$ (0.1 μW to 1 mW), a) output DC power and b) overall efficiency, at 680 MHz.

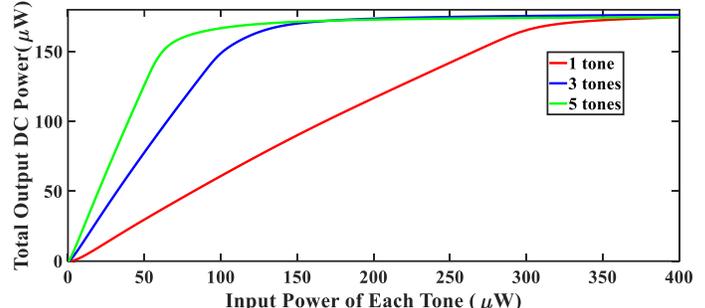

Fig. 14. Simulated results of the output DC power as a function of input power for the proposed MSNC in power saving mode with multi-tone excitations (1 tones, 3 tones, and 5 tones) over input power range of 0 to 400 $\mu W$ (-4 dBm).

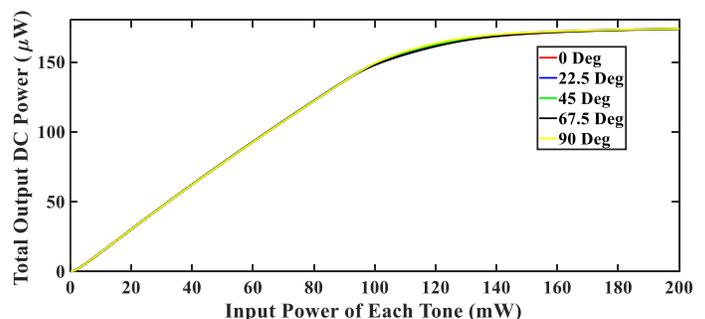

Fig. 15. Simulated phase mismatch (0°, 22.5°, 45°, 67.5°, and 90°) between 3-tone excitation signals.

TABLE IV
COMPARISON OF THE PROPOSED MSNC SYSTEM AND RELATED REFERENCES

| Ref. | Frequency (GHz) | Max. Efficiency at Low Power Range | Function |
|---|---|---|---|
| [31] | 0.95 | 60% @ -12 dBm | Single-band rectifier |
| [32] | 1.8 | 55% @ +10 dBm | Wideband rectifier |
| [33] | 0.9, 2.4 | 70% @ +15 dBm | Dual-band rectifier |
| [34] | 0.85, 1.77, 2.07 | 60% @ 0 dBm | Triple band rectifier |
| [35] | 0.75, 1.8, 2.4, 5.8 | 70% @ -10dBm | Quad-band rectifier |
| [36] | 2.45 | 45% @ -10 dBm | 2-mode system: EH and WPT system |
| This Work | 0.68 | 60% @ -12 dBm | 3- mode system: Rx, Tx and power saving modes |
9

## IV. Conclusion

A new multimode smart nonlinear circuit (MSNC), which is sensitive to the input power level has been presented in this paper. The proposed MSNC operates in wireless communications (Tx and Rx modes) as well as energy harvesting (EH) and power saving modes. Low cost, compactness, high sensitivity, and integrability make the proposed MSNC a good candidate for IoT devices. The measured, simulation and theoretical results have been presented to verify the functionality of the proposed MSNC. Based on the achieved results, it is evident that the proposed MSNC exhibits better than 13 dB return loss at 680 MHz for the desired power range, over 60% efficiency at port 3 (power saving mode), and less than 1 dB transmission loss at RF ports, which are port 1(Ant) and port 2 (TR). It should be mentioned that the measured and simulation results are in good agreement, which verifies the design procedure accuracy. Furthermore, the proposed MSNC is insensitive to the phase mismatch for multi-tone excitation and is a good candidate for real WPT scenarios. Further, as the future work, an antenna and supercapacitor/battery would be included to realize a smart switching system, as an alternative green energy source for integrated communication and sensing applications. Moreover, a power management unit (e.g. LT3107, ADP5092) can be connected between MSNC and power storage unit (supercapacitor/battery) to control the load impedance and charging procedure, properly, and it presents a constant input impedance.

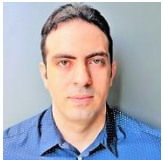

**Rasool Keshavarz** was born in Shiraz, Iran in 1986. He received the Ph.D. degree in Telecommunications Engineering from the Amirkabir University of Technology, Tehran, Iran in 2017 and is currently working as Senior Research Fellow in RFCT Lab at the University of Technology, Sydney, Australia. His main research interests are RF and microwave circuit and system design, sensors, antenna design, wireless power transfer (WPT), and RF energy harvesting (EH).

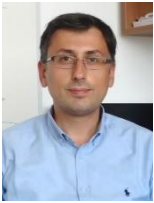

**Dimitrios Sounas** is an Assistant Professor at the ECE Department at Wayne State University. He has received his Ph.D. degree in Electrical and Computer Engineering from the Aristotle University of Thessaloniki, Greece, in 2009. From 2010-2018 he has been a postdoctoral fellow and a research scientist at Polytechnique Montreal and the University of Texas at Austin.

He has made major contributions in magnet-less nonreciprocal devices with applications in full-duplex systems and radars. He has contributed to the founding of Silicon Audio RF Circulator, Austin, TX, USA, specializing in the design of angular-momentum circulators for RF and acoustical systems. He has authored or co-authored over 100 papers in peer-reviewed journals, including Science and various Nature journals. He has more than 150 conference abstracts, 6 patents, and more than 8,000 citations. His current research focuses on time-modulated and topological metamaterials, optical signal processing, and inverse design, and is funded by the Department of Defense.

Dimitrios is a Senior Member of IEEE and the recipient of the 2020 EurAAP Leopold B. Felsen Award for Excellence in Electrodynamics. He has chaired and organized various sessions in international symposia. He is an Associate Editor for the IEEE Transactions on Antennas and Propagation and a reviewer for more than 20 engineering and physics journals.

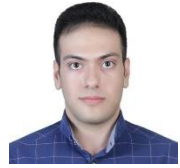

**Saeed Keshavarz** received his M.Sc. from Amirkabir University of Technology (AUT) in 2018, and he is currently PhD student at Wayne State University (WSU) in Electrical Engineering. His research interests contain passive and active microwave devices using composite right and left-handed (CRLH) transmission lines and magnetless circulators based on spatiotemporal modulation.

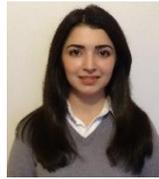

**Negin Shariati** is an Associate Professor in the School of Electrical and Data Engineering, Faculty of Engineering and IT, University of Technology Sydney (UTS), Australia. She established the state of the art RF and Communication Technologies (RFCT) research laboratory at UTS in 2018, where she is currently the Co-Director and leads research and development in RF Technologies, Sustainable Sensing, Energy Harvesting, Low-power Internet of Things and AgTech. She leads the Sensing Innovations Constellation at Food Agility CRC (Corporative Research Centre), enabling new innovations in agriculture technologies by focusing on three key interrelated streams; Sensing, Energy and Connectivity. Since 2018, she has held a joint appointment as a Senior Lecturer at Hokkaido University, externally engaging with research and teaching activities in Japan. Negin Shariati was the recipient of Standout Research Award in the 2021 IoT Awards Australia, and IEEE Victorian Section Best Research Paper Award 2015. She attracted over two million dollars worth of research funding across a number of CRC and industry projects, where she has taken the lead CI (Chief Investigator) role and also contributed as a member of the CI team. Dr Shariati completed her PhD in Electrical-Electronics and Communication Technologies at Royal Melbourne Institute of Technology (RMIT), Australia, in 2016. She worked in industry as an Electrical-Electronics Engineer from 2009-2012. Her research interests are in RF-Electronics Circuits and Systems, Sensors, Antennas, RF Energy Harvesting, Simultaneous Wireless Information and Power Transfer, and Wireless Sensor Networks.